# A COOPERATIVE CONTROL FRAMEWORK FOR CAV LANE CHANGE IN A MIXED TRAFFIC ENVIRONMENT

**Runjia Du**
Graduate Research Assistant, Center for Connected and Automated Transportation (CCAT), and Lyles School of Civil Engineering, Purdue University, West Lafayette, IN, 47907.
Email: du187@purdue.edu
ORCID #: 0000-0001-8403-4715

**Sikai Chen***
Postdoctoral Research Fellow, Center for Connected and Automated Transportation (CCAT), and Lyles School of Civil Engineering, Purdue University, West Lafayette, IN, 47907.
Email: chen1670@purdue.edu; and
Visiting Research Fellow, Robotics Institute, School of Computer Science, Carnegie Mellon University, Pittsburgh, PA, 15213.
Email: sikaichen@cmu.edu
ORCID #: 0000-0002-5931-5619
(Corresponding author)

**Yujie Li**
Graduate Research Assistant, Center for Connected and Automated Transportation (CCAT), and Lyles School of Civil Engineering, Purdue University, West Lafayette, IN, 47907.
Email: li2804@purdue.edu
ORCID #: 0000-0002-0656-4603

**Jiqian Dong**
Graduate Research Assistant, Center for Connected and Automated Transportation (CCAT), and Lyles School of Civil Engineering, Purdue University, West Lafayette, IN, 47907.
Email: dong282@purdue.edu
ORCID #: 0000-0002-2924-5728

**Paul (Young Joun) Ha**
Graduate Research Assistant, Center for Connected and Automated Transportation (CCAT), and Lyles School of Civil Engineering, Purdue University, West Lafayette, IN, 47907.
Email: ha55@purdue.edu
ORCID #: 0000-0002-8511-8010

**Samuel Labi**
Professor, Center for Connected and Automated Transportation (CCAT), and Lyles School of Civil Engineering, Purdue University, West Lafayette, IN, 47907.
Email: labi@purdue.edu
ORCID #: 0000-0001-9830-2071

**ABSTRACT**

In preparing for connected and autonomous vehicles (CAVs), a worrisome aspect is the transition era which will be characterized by mixed traffic (where CAVs and human-driven vehicles (HDVs) share the roadway). Consistent with expectations that CAVs will improve road safety, on-road CAVs may adopt rather conservative control policies, and this will likely cause HDVs to unduly exploit CAV conservativeness by driving in ways that imperil safety. A context of this situation is lane-changing by the CAV. Without cooperation from other vehicles in the traffic stream, it can be extremely unsafe for the CAV to change lanes under dense, high-speed traffic conditions. The cooperation of neighboring vehicles is indispensable. To address this issue, this paper develops a control framework where connected HDVs and CAV can cooperate to facilitate safe and efficient lane changing by the CAV. Throughout the lane-change process, the safety of not only the CAV but also of all neighboring vehicles, is ensured through a collision avoidance mechanism in the control framework. The overall traffic flow efficiency is analyzed in terms of the ambient level of CHDV-CAV cooperation. The analysis outcomes are including the CAVs lane-change feasibility, the overall duration of the lane change. Lane change is a major source of traffic disturbance at multi-lane highways that impair their traffic flow efficiency. In providing a control framework for lane change in mixed traffic, this study shows how CHDV-CAV cooperation could help enhancing system efficiency.

**Keywords:** autonomous vehicle, connectivity, cooperative level, relative velocity.

## I. INTRODUCTION

### Safety in the mixed flow era

Recent development in Autonomous Vehicle (AV) capabilities have been discussed as a collective set of disruptive technologies that will profoundly impact the current transportation system in terms of safety. Traffic-related accidents are the second leading cause of death between the age of 5 and 29, and the third leading cause of death between the age of 30 and 44 (Anjuman et al., 2020). According to a report from the National Highway Traffic Safety Administration, the potential of AVs to save lives and reduce injuries is particularly welcome due to the realization that 94% of serious crashes are due to human error. Autonomous driving is expected to remove or at least mitigate human error and thereby reduce crashes (NHTSA; Chen, 2019; Chen et al., 2020). This can be most effective when the market penetration of the AVs is 100%. However, the transition phase of HDVs to AVs will last longer than most stakeholders expected (Hancock et al., 2019). For example, the Volvo CEO Hakan Samuelsson recently said it was "irresponsible" to put AVs on the road if they were not sufficiently safe, because that would erode trust among the public and regulators (Kristy, 2019). Consistent with his statement, it can be argued that on-road AVs should be "sufficiently" safe, but this could cause the AV to behave too conservatively. Such concern is well justified, because on-road CAVs may adopt such conservative control policies as they are expected to improve road safety, and this will likely cause HDVs to unduly exploit the CAV conservativeness by driving in ways that imperil safety. In other words, realizing the tolerant nature of CAV operations, HDVs may undertake driving behaviors (for example, texting) and maneuvers (for example, keeping small headways) that are inherent less safe than they would do in an all-HDV traffic environment (Li et al., 2020a). Such HDV behavior may be deliberate and conscious, or may be due to risk compensation. A specific context of this situation is lane changing by the CAV, as we discuss in the next section of this paper.

### AV Lane change challenge

The lane-changing maneuver is critical to road safety as 40% of freeway accidents occur in ramp areas (Chen et al., 2009). When traffic on the fast lane need to exit the freeway, they will need to perform multiple lane changes, which is extremely unsafe for HDVs and much more so for AVs. As AVs lack human dexterity (at least in their current stage of development), they need to be conservative so that their operations can be safe. As we stated in the previous section, conservative behavior not only leaves AVs vulnerable to aggressive human drivers and inhibits the interpretability of intentions. In a recent analysis of California traffic incidents with AVs, in 57% of crashes, the AV was rear-ended by human drivers. In real-world traffic, HDVs (human-driven vehicles) commonly violate regulations, but AVs are likely to be required to obey all rules and regulations. Therefore, AVs may be "bullied" by HDVs. One such example is Google's self-driving car, which experienced a crash because an HDV violated a stop sign (Stewart, 2015). Figure 1 presents two examples where the AV needs to perform mandatory lane changes. Where there is dense traffic on the target lane, it is difficult for an AV to change lanes without the cooperation of other human-driven vehicles. Further, lane-changing maneuvers are a major cause of traffic disturbance on multi-lane freeways (Barria and Thajchayapong, 2011; Ha et al., 2020a). Low efficiency lane-changing maneuvers will delay the traffic near the ramp (Gong, 2016; Zheng, 2019; Atagoziyev, 2016; Dong et al., 2020a). It can be hypothesized that AV-HDV cooperation (through connectivity) can enhance the efficiency of the overall traffic flow.

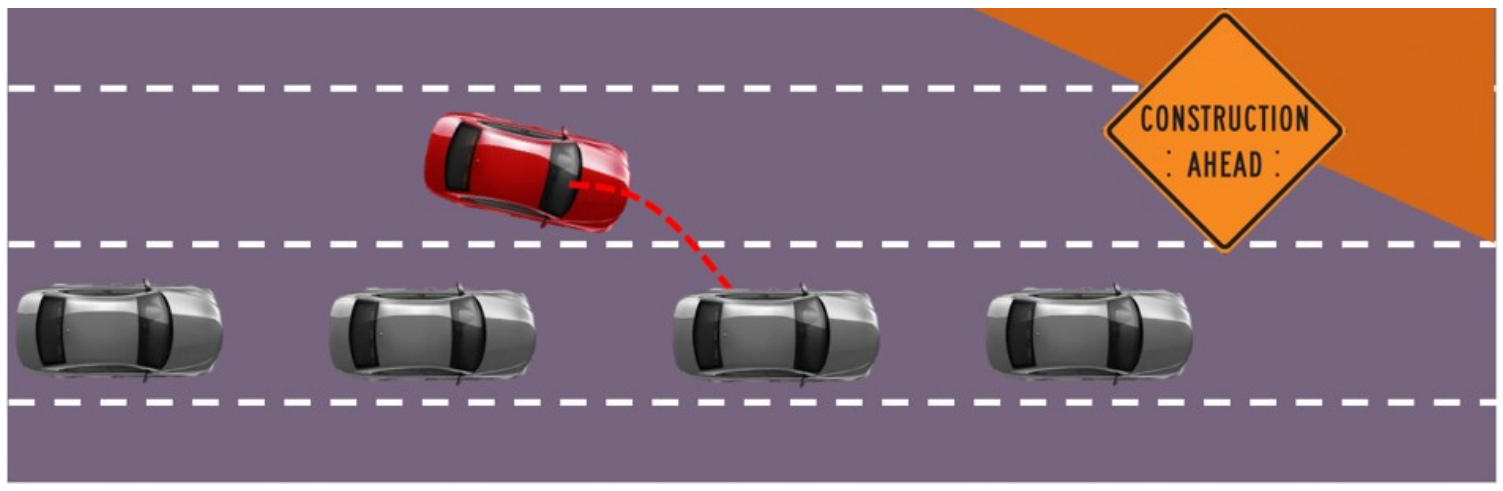

(a) AV lane-changing to avoid the construction area ahead.

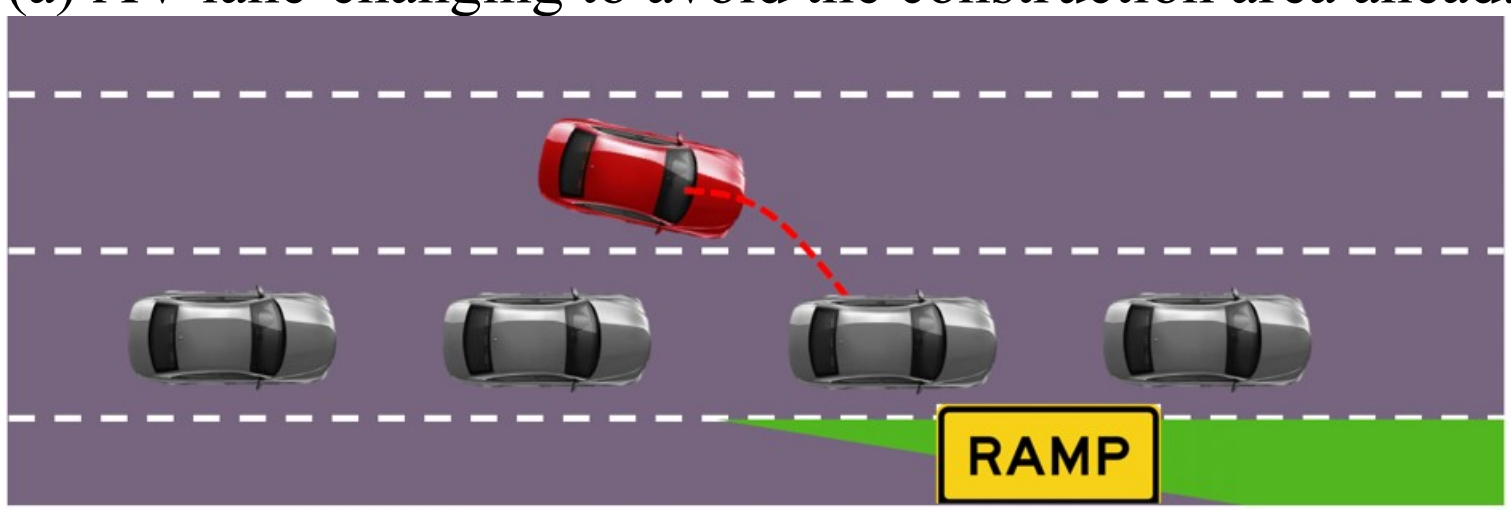

(b) AV lane-changing to exit the ramp.

**FIGURE 1. Mandatory Lane-changing Maneuvers**

**Connectivity can enhance safety**

Connectivity technology also plays a role in magnifying AV capabilities and benefits. Through connectivity, vehicles will be able to communicate with other vehicles. In situations involving hazardous roadway conditions, drivers of CHDVs can receive warning notifications and alerts. NHTSA duly recognizes that connectivity is a promising technology that can prevent 615,000 crashes and thereby significantly reduce the number of fatalities and serious injuries associated with highway crashes (NHTSA, 2019). Further, a core aspect of the USDOT's intelligent transportation system research program is the Connected Vehicle Research Initiative (USDOT, 2011). Connectivity technology is less expensive compared to sensor technologies, and this makes it more affordable and practical for installation on AVs and HDVs. With the help of connectivity, the vehicles' cooperation can be enhanced (Dong et al., 2020b; Li et al., 2020b). Therefore, the lane-changing maneuver of AV in the mixed traffic flow can be facilitated through cooperation with the neighboring vehicles in the traffic environment, and this could address one of the many challenges that AVs are expected to encounter in the prospective era of HDV-AV transition (Ha et al., 2020b).

**Cooperative control by MPC**

With connectivity, the cooperative control can be realized though temporarily speed/control override of the surrounding vehicles by the centralized control platform, which can be considered as the extension of the Cooperative Adaptive Cruise Control (Milanés, 2013). In a complex environment of multiple vehicles interact with each other, the decentralized control decisions will be based on incomplete information. However, the centralized control that combine the global information in the system may yield better performance (Blunck et al., 2018). Moreover, Model Predictive Control (MPC) is effective for solving motion planning problems. The MPC method formulated the motion planning problem as an optimization problem, often as a constrained, convex problem solved in a recursive manner by considering the updating of the environment states during the planning process (Liu, 2017). MPC control method is widely used to solve motion planning problems in the literatures (Babu et al., 2019; Wang et al., 2019; Du et al., 2020; Shen et al., 2017; Werling et al., 2012; Ji et al., 2016)

This study focuses on formulating a control framework that combines connectivity and automation (CHDVs + CAV) to help CAVs in mixed traffic to successfully change lanes. Also, the outcomes of the control framework, in terms of safety and efficiency, are considered. Safety benefits are expected to pertain not only to the CAV and nearest CHDVs, but also other vehicles in the outer neighborhood of the lane-change location. The framework does not do away with human drivers. CHDVs refer to human-driven vehicles that have connectivity capability and therefore communicate with the CAV and with each other. Therefore, this study considers the cooperative intention of the human drivers as an important factor. In this study, we consider several different combinations of the vehicle cooperative intention levels and locations. The cooperative control framework is proposed using the Model Predictive Control (MPC) theory. This paper is organized as follows: Section II describes the problem formulation of the cooperative control framework. Section III introduces the detailed mathematical models and controller design. Section IV presents and analyzes the simulation results. Section V summarizes the study and discusses the study limitations and directions for future work.

## II. PROBLEM FORMULATION

Consider a CAV on a highway segment that need to change lanes from current lane to a target lane. The reason for lane change could be a construction work zone or an accident located on the same lane downstream. If the CHDVs on the target lane cooperate with the CAV (through communication), they will create a gap so that the CAV can change lanes successfully. In practice, the CHDVs are still human-driven, which means that the human drivers are still involved in the system. Thus, the cooperative intention of the CHDVs will not always be 100%. The effect of different levels of cooperative intention, as well as the different locations, need to be considered in the control framework. Through the lane-changing process, the control framework needs to guarantee a certain minimum safety requirements in terms of relative distance and relative velocity between any CAV-CHDV pair. Throughout the lane-change process, the relative distance is determined based on the relative velocities between the CAV and the CHDV and therefore, will influence the decisions made by the CAV's control framework. This framework will control the acceleration/deceleration rate of the CHDVs on the target lane based on the safety requirements and efficiency. The problem settings and control framework are introduced in the following subsections.

### Problem settings

As stated above, the control framework considers both the cooperative intention and relative position of the CHDVs on the target lane. Figure 2 illustrates an example of the combination of cooperative intention levels and relative positions. The CHDVs are shown at different locations and cooperation levels. The green vehicles are those which actively cooperate, while the grey vehicles are those that are less active in cooperating. The vehicles within the blue square are those located "near" the CAV; and the vehicles in the yellow square but not in the blue square is those located "far" from the CAV.

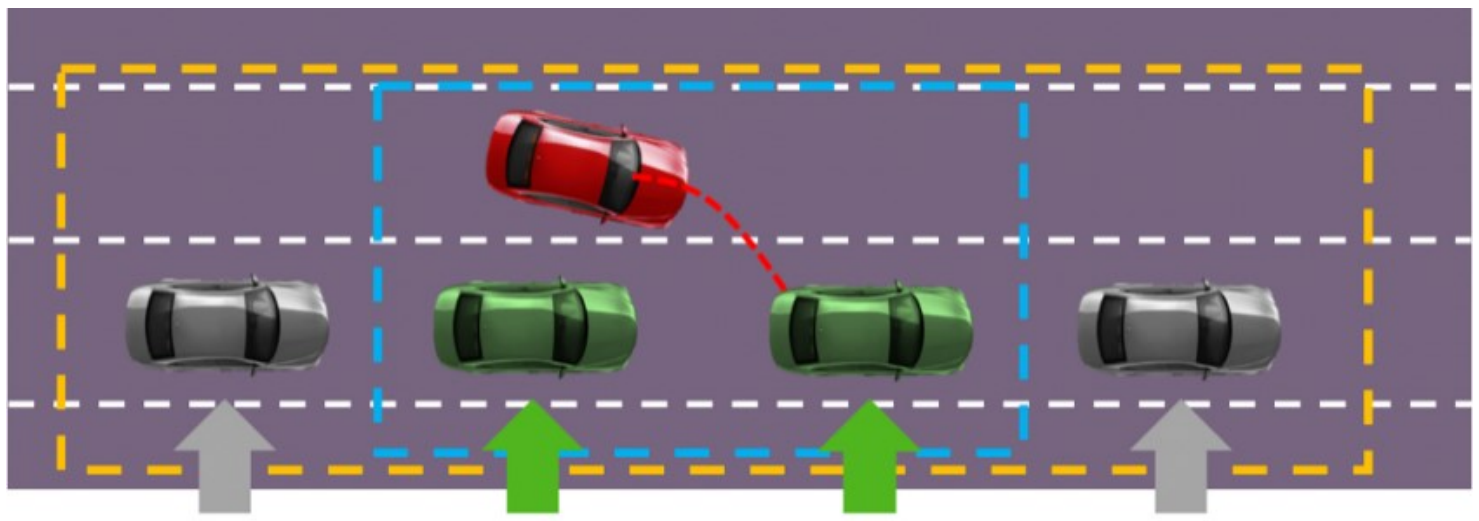

**FIGURE 2. CHDVs with different cooperation levels and locations**

The CHDVs on the target lane are concerned about the vehicles that neighbor them. Thus, the CHDVs in the "near" location need to consider both the CAV and CHDVs in their immediate vicinity. The CHDVs in the "far" location consider only the CHDV in the "near" location. Moreover, based on the position of the CAV, the CHDVs that are longitudinally preceding the CAV are termed PHDV (Preceding human-driven vehicle); and the CHDVs that are longitudinally following the CAV are termed FHDV (Following human-driven vehicle). Regarding the target CAV, the planned lane-changing trajectory needs to consider the CHDVs in the "near" location. The ending position of the lane-changing trajectory needs to be within the future ending positions of the CHDVs in the "near" location, which is greater than the ending position of FHDV and smaller than the ending position of PHDV.

**Assumptions**

This study is based on the following assumptions:

(1) The initial bumper-to-bumper headway is set based on the initial velocity.
(2) The planned trajectory in each time step, is a cubic polynomial curve (Yang et al., 2018).
(3) The CHDVs on the target lane do not perform lane-change maneuvers throughout the CAV's lane change process.
(4) The CAV's lane changing motion is planned based on the future end position of the FHDV and PHDV in the "near" location.
(5) The velocities of all the vehicles considered in the system are normally distributed.

According to our previous research, using a buffer area to represent the vehicle will enhance the safety considerations (Wang and Schwager, 2019; Shaffer and Herb, 1992). In this research, we use circle buffer area. In order to avoid a collision, the buffer circles need to avoid intersection or even tangential situations. Figure 3 presents the buffer area and the situations that need to be avoided.

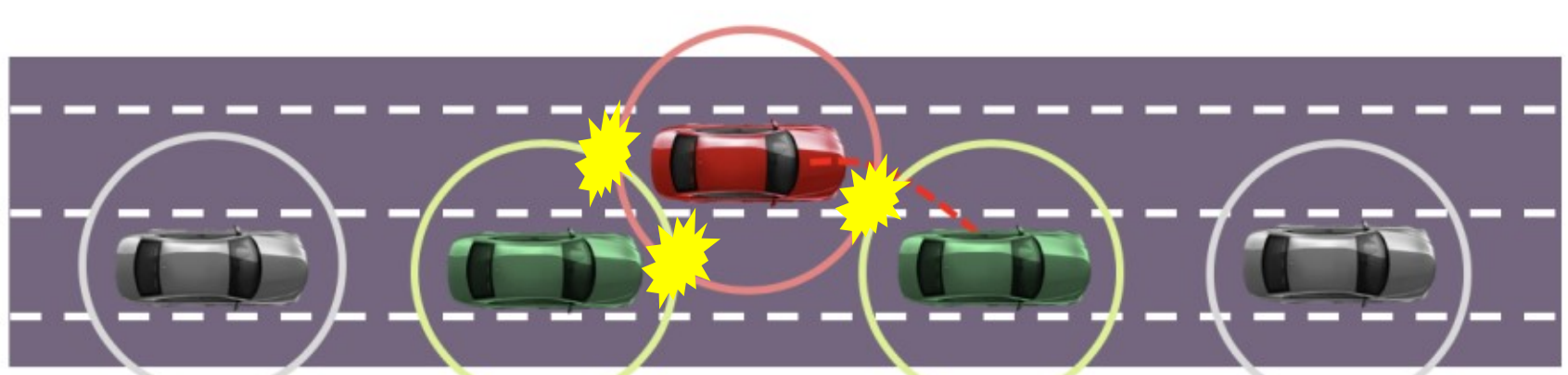

**FIGURE 3. Buffer area and situations need to be avoided (intersect, tangent)**

## III. METHODOLOGY

In this section, the cooperative control framework that based on the MPC method, is introduced, and the detailed CAV lane-changing trajectory planning model and motion models of the controlled CHDVs are formulated. Further, the optimization problem of the MPC control framework is discussed in detail.

**Control framework**

The design of cooperative control framework is to assist the CAV's lane-changing process. It is important to consider the lane-changing motion of both the CAV and the CHDVs. The lane-changing trajectory planning model is formulated from two considerations: (a) lane-changing trajectory, (b) Velocity. The control framework uses Model Predictive Control (MPC) as its basis, and considers current states of CHDVs (longitudinal positions, velocity) as the inputs and the acceleration/deceleration as output. Since the problem has several complicated constraints under different conditions, the MPC controller is used to deal with multiple constraints (Pisaturo et al., 2014; Liu et al., 2014). For the optimal problem in the MPC controller, the CAV's lane-changing trajectory and velocity are used as reference for the CHDVs to avoid sensitive control, and large control efforts are penalized. Figure 4 presents the block diagram of the cooperative control framework.

Moreover, the cooperative intention level is considered by the soft constraints of velocity. For the actively cooperate CHDVs, the velocity will be strictly satisfying the safety requirements. However, for those CHDVs who are less active in cooperating the target CAV, there will be soft constraints on the velocity, which means the velocity can violate by certain level through the lane-changing process. The detailed controller design will be introduced in the following subsections.

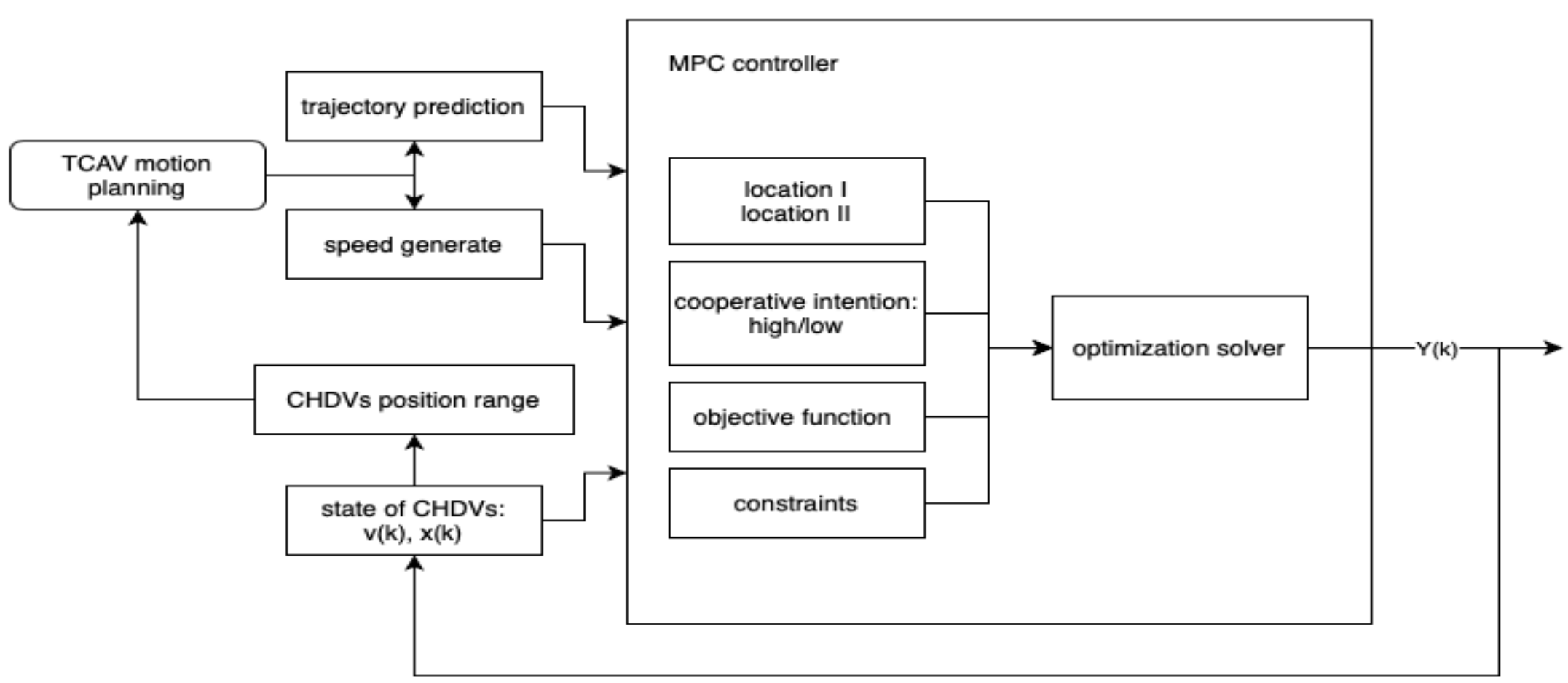

**FIGURE 4. The block diagram of the cooperative control framework**

**Target CAV trajectory planning model**

Motion planning consist of (a) trajectory planning, (b) speed profile generation. The planned trajectory at each time step is represented by a cubic polynomial curve ($y(x) = ax + bx^2 + cx^3$), which has the second smoothness (Yang et al., 2018).

$$y_t(x_t) = -tan\theta_t^i x_t + \frac{2x_t^e tan\theta_t^i - 3y_t^t}{(x_t^e)^2} x_t^2 + \frac{2y_t^e - x_t^e tan\theta_t^i}{(x_t^e)^3} x_t^3 \quad (1)$$

In Equation (1), $(x_t, y_t)$ is the position of the target CAV at time $t$. $x_t$ and $y_t$ is the longitudinal and latitudinal positions of the TCAV (Target CAV) respectively. $x_t^e$ represents the ending position of the Target CAV, which decided by the rollover-free boundary as well as the future ending positions of the CHDVs (discussed in a subsequent section of this paper) in the near location of the TCAV. $\theta$ represents the course angle of the LHDV, which is the angle between the moving direction $U$ and the $x-axis$. $\theta_t^i$ is the initial moving direction angle of time step $t$. The final position's course angle ($\theta_e = 0$ ;$y_t'(x_t^e)$=0). The rollover-free boundary is calculated based on Wang et, al's model:

$$x_n^f = \frac{6y_t^e u_t^i}{\sqrt{6y_t^e a_s^r}} \tag{2}$$

$y_t^e$ represents the ending latitudinal position, and $u_t^i$ represents the initial velocity towards moving direction for time step $t$. The $a_s^r$ is the boundary latitudinal acceleration, and the value is set as $6.958\,m/s^2$, which is based on Sun and Wang's research regarding lane-changing lateral acceleration (Sun and Wang, 2018).

The speed profile is generated with the purpose of completing the lane-changing in very short time as possible. The velocity profile of the TCAV can be calculated based on the ending position. When the rollover-free boundary is considered an appropriate ending position, the proper longitudinal acceleration can be represented as:

$$a = \frac{2\left(\frac{6y_t^e u_t^i}{\sqrt{6y_t^e a_s^r}} - u_t^i \tau\right)}{\tau^2} \tag{3}$$

$\tau$ represents the length of a single time step. If we assume the highest longitudinal acceleration rate is $a_{max}$, then the proper aggressive longitudinal acceleration can be represented as $min\{a, a_{max}\}$. Based on the acceleration, the aggressive velocity of the time step $t$ is therefore $u_n^i + \min\{a, a_{max}\}\tau$.

However, the TCAV future ending position also needs to consider the ending positions of the CHDVs in the near location. Therefore, to ensure safety, the final longitudinal position of the planned lane-changing trajectory needs to consider both the future ending positions of the CHDVs and also the rollover-free condition. At each time step, the planned lane-changing trajectory needs to compare the rollover-free ending positions: $(x_f^r, \infty)$ and the ending positions of the CHDVs, which are FHDV and PHDV in the "near" location: $(e_{FHDV1}, e_{PHDV1})$. When a proper final position is decided, the moving direction angle and longitudinal distance of each time step will change based on the final position. Figure 5 presents the detailed trajectory planning process:

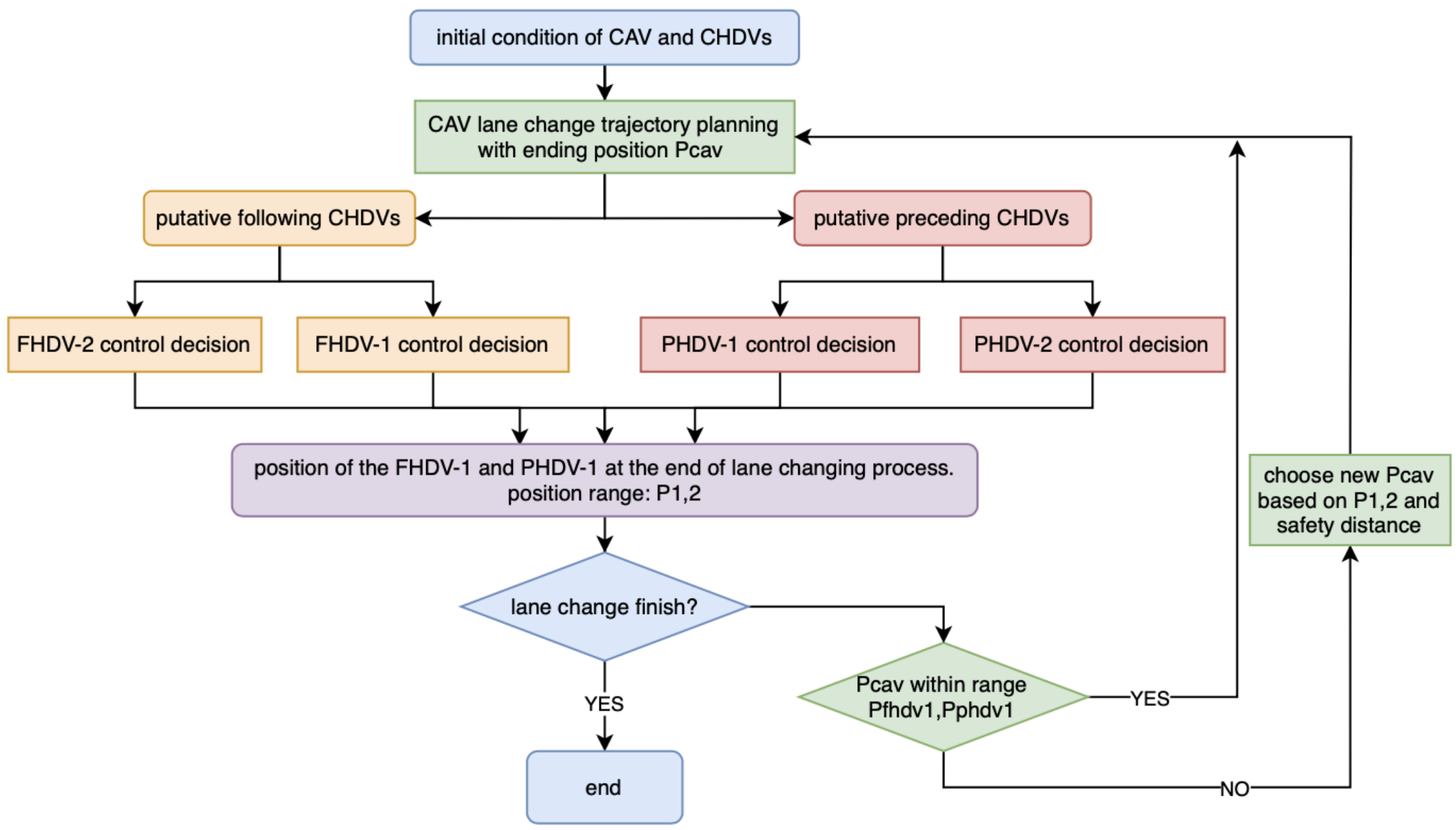

**FIGURE 5. Flow chart of the CAV's trajectory planning**

**Controlled CHDVs motion model**

The CHDVs in this research refer to only those HDVs with connectivity. The controller can control only their longitudinal acceleration/deceleration and velocity but not their lane-changing maneuvers. The motion model for each time step can be represented as a discrete-time model:

$$x(k+1) = x(k) + v(k)\tau + \frac{1}{2}a(k)\tau^2 \quad (4\text{-}1)$$
$$y(k+1) = y(k) = c \quad (4\text{-}2)$$
$$v(k+1) = v(k) + a\tau \quad (4\text{-}3)$$

$c$ represents the constant latitudinal positions of the CHDVs. And the motion model can be written more compactly as:

$$\bar{x}(k+1) = A\bar{x}(k) + Bu(k) \quad (5\text{-}1)$$
$$y(k+1) = C\bar{x}(k) \quad (5\text{-}2)$$
$$for: \bar{x}(k) = \begin{bmatrix} x(k) \\ v(k) \end{bmatrix}, A = \begin{bmatrix} 1 & \tau \\ 0 & 1 \end{bmatrix}, B = \begin{bmatrix} \frac{1}{2}\tau \\ \tau \end{bmatrix} \quad (5\text{-}3)$$

With $C$ equals to $I$, the MPC controller can predict the states of the system in the multiple sampling times to have a more accurate decision. $N_p$ represents the prediction horizon, which is the number of future control intervals that MPC evaluates. $N_c$ represents the control horizon, which is the number of control actions to optimize the control interval. In this study, the relation between $N_p$ and $N_c$ is to be: $N_p = N_c + 1$. Thus, the predicted outputs through the control interval can be written as:

$$\bar{x}(k+1) = A\bar{x}(k) + Bu(k) \quad (6\text{-}1)$$
$$\bar{x}(k+2) = A^2\bar{x}(k) + ABu(k) + Bu(k+1) \quad (6\text{-}2)$$

…

$$\bar{x}(k+N_c) = A^{N_c}\bar{x}(k) + ABu(k) + \cdots + Bu(k+N_c-1) \tag{6-3}$$

$$\bar{x}\left(k+N_p\right) = A^{N_p}\bar{x}(k) + ABu(k) + \cdots + (AB+B)u(k+N_c-1) \tag{6-4}$$

Thus, the output sequence and input sequence can be represented in a more compact form as:

$$X(k+1) = \begin{bmatrix} \bar{x}(k+1) \\ \bar{x}(k+2) \\ \bar{x}(k+3) \\ \vdots \\ \bar{x}\left(k+N_p\right) \end{bmatrix}_{N_p\times 1} \qquad U(t_i) = \begin{bmatrix} u(k) \\ u(k+1) \\ u(k+2) \\ \vdots \\ u(k+N_c-1) \end{bmatrix}_{N_c\times 1} \tag{7}$$

Based on the output and input sequence, the system prediction can be written as: $X(k+1) = M_x\bar{x}(k) + M_uU(k)$, where:

$$M_x = \begin{bmatrix} A \\ A^2 \\ \vdots \\ A^{N_p} \end{bmatrix}_{N_p\times 2} \quad M_u = \begin{bmatrix} B & 0 & \cdots & 0 \\ AB & B & \cdots & 0 \\ \vdots & \vdots & \ddots & \vdots \\ A^{N_c-1}B & A^{N_c-2}B & \cdots & B \\ A^{N_p-1}B & \cdots & \cdots & (A+1)B \end{bmatrix}_{2N_p\times N_c} \tag{8}$$

**Optimization problem of MPC controller**

The MPC controller considers the relative position of TCAV and other CHDVs. Based on the initial positions, there are two types of locations that are shown in Figure 2. For the CHDVs in "near" location, the interactions considered are TCAV and neighboring CHDVs. For the CHDVs in the "far" location, the interactions considered are the CHDVs around. The objective is to control the CHDVs and TCAV to cooperatively complete the lane-changing process while avoiding collision of any vehicles in the system. If the feasibility cannot be satisfied at the first try, then the CHDVs will be controlled to adjust their velocities to create the gap and have the lane-changing process start feasibly. The large control efforts are penalized in the optimization problems. The objective functions are formulated differently in terms of the location of the CHDVs. Moreover, the constraints will be different for the following vehicles and preceding vehicles.

CHDVs in the "near" location: they consider both the TCAV and the neighboring CHDV in "far" location. The objective function consists of the tracking of the TCAV, the control inputs and velocity soft constraints.

$$J_1 = \min_{\bar{u},\bar{\delta}} \sum_{k=1}^{N_p}\|\bar{x}(k+n) - r(k+n)\|_Q^2 + \sum_{k=1}^{N_c}\|\bar{u}(k+n-1)\|_R^2 + \left\|\bar{\delta}(k+n-1)\right\|_P^2 + Jerk \tag{9}$$

$s.t. (n = 1, \ldots, N_p)$

$$\bar{x}(k+n) = A\bar{x}(k+n-1) + B\bar{u}(k+n-1) \tag{10}$$

$$d_{max} \le \bar{u}(k+n-1) \le a_{max} \tag{11}$$

Equation (10) represents the system equation, and the $d_{max}$, $a_{max}$ in Equation (11) represent the maximum deceleration and acceleration rates. Moreover, the comfort (lack of jerk) is considered. CHDVs give up their speed to cooperate with the TCAV, and in practice, the comfort (or, lack of jerk) that is associated with the deceleration/acceleration, is an important factor that influences the CHDV driver's compliance with the CAV controller recommendations. Jerk has a significant impact on the safety and comfort of passengers. The jerk cost function considers the change of the acceleration/deceleration through the control horizon:

$$J = \int_{t_0}^{t_f} \sum_{i=1}^{m} \tilde{u_i}^2 \, dt \tag{12}$$

In this research, we consider using the prediction horizon as $N_p = 5$, control horizon as $N_p = N_c - 1 = 4$, which are based on previous research. The difference between the control inputs can be represents in the compact form:

$$[u_1, u_2, u_3, u_4]^T C \begin{bmatrix} u_1 \\ u_2 \\ u_3 \\ u_4 \end{bmatrix} \tag{12-1}$$

$$C = \begin{bmatrix} 1 & -1 & 0 & 0 \\ -1 & 2 & -1 & 0 \\ 0 & -1 & 2 & -1 \\ 0 & 0 & -1 & 1 \end{bmatrix} \tag{12-2}$$

For the CHDV that is putative following the TCAV on the target lane, the constraints can be described as follows:

$$\bar{x}(k+n) - r(k+n) \le \begin{bmatrix} -l_1 \\ \bar{\delta}(k+n-1) \end{bmatrix} \; for\!: k = 1, \dots, N_c \tag{13-1}$$

$$\bar{x}(k+n) - r(k+n) \le \begin{bmatrix} -l_1 \\ 0 \end{bmatrix} \; for\!: k = N_p \tag{13-2}$$

$$\bar{x}(k+n) - r(k+n) \ge \begin{bmatrix} G\sqrt{4R^2 - \left(l_y^{\bar{x}}(k+n) - l_y^{r}(k+n)\right)^2} \\ -\bar{\delta}(k+n-1) \end{bmatrix} \tag{13-3}$$

$$\bar{u}(k+n-1) \ge \frac{2\left(l_2 - p\left(\Delta x(k+n-1)\right)\right)}{\tau^2} + d_{max} \tag{13-4}$$

For the CHDV that is putative preceding to the TCAV on the target lane, the constraints are different:

$$\bar{x}(k+n) - r(k+n) \ge \begin{bmatrix} l_1 \\ -\bar{\delta}(k+n-1) \end{bmatrix} \; for\!: k = 1, \dots, N_c \tag{14-1}$$

$$\bar{x}(k+n) - r(k+n) \ge \begin{bmatrix} l_1 \\ 0 \end{bmatrix} \; for\!: k = N_p \tag{14-2}$$

$$r(k+n) - \bar{x}(k+n) \ge \begin{bmatrix} G\sqrt{4R^2 - \left(l_y^{\bar{x}}(k+n) - l_y^{r}(k+n)\right)^2} \\ \bar{\delta}(k+n-1) \end{bmatrix} \tag{14-3}$$

$$\bar{u}(k+n-1) \le a_{max} + \frac{2(p(\Delta x(k+n-1)) - l_2)}{\tau^2} \quad (14\text{-}4)$$

The $r(k+n)$ represents the state of TCAV, which includes the longitudinal location and the velocity. $\Delta x(k+n-1)$ in Equations (13-4) and (14-4) represents the state difference between the CHDV in "near" location ($\bar{x}(k+n-1)$) and CHDV in "far" location ($\hat{x}(k+n-1)$). Thus, $\Delta x(k+n-1)$ equals to $\bar{x}(k+n-1) - \hat{x}(k+n-1)$. The TCAV tracking in the objective function penalizes the aggressive control action. Moreover, the large control efforts will be penalized. The cooperative intention is shown through the $\bar{\delta}(k+n-1)$, which denotes the violation allowance of the bound in the velocity constraint. Furthermore, the distance constraints (13-1), (13-2), (14-1), (14-2) are applied to guarantee the safe requirements of longitudinal distance. $l_1$ represents the safety requirements in terms of longitudinal distance between TCAV and CHDVs in the "near" location. $l_2$ represents the safety requirements of longitudinal distance between CHDVs in the "near" location and "far" location. Also, the crash avoidance constraints (13-3), (14-3) are applied to make sure the safety by avoiding the tangent situations of the circle buffer areas. Constraints (13-4), (14-4) are used to illustrate the relation between the CHDVs on the target lane that are in the different locations. Those constraints help avoid any imminent secondary collision between CHDVs that neighbor each other.

CHDVs in the "far" location: These consider the CHDVs that neighbor them, which are the CHDVs in the "near" location. The objective function consists of the tracking of the CHDV from "near" location, control inputs and the soft constraints for the velocities.

$$J_2 = \min_{\hat{u},\hat{\delta}} \sum_{k=1}^{N_p} \|\hat{x}(k+n) - \bar{x}(k+n)\|_Q^2 + \sum_{k=1}^{N_c} \|\hat{u}(k+n-1)\|_R^2 + \left\|\hat{\delta}(k+n-1)\right\|_P^2 + Jerk \quad (15)$$

$$s.t.\,(n = 1, \ldots, N_p)$$

$$\hat{x}(k+n) = A\hat{x}(k+n-1) + B\hat{u}(k+n-1) \quad (16)$$

$$d_{max} \le \hat{u}(k+n-1) \le a_{max} \quad (17)$$

For the CHDVs in "far" location in the following position of the CHDV in "near" location on the target lane, the constraint set is:

$$\hat{x}(k+n) - \bar{x}(k+n) \le \begin{bmatrix} -l_2 \\ \bar{\delta}(k+n-1) \end{bmatrix} for\!: k = 1, \ldots, N_c \quad (18\text{-}1)$$

$$\hat{x}(k+n) - \bar{x}(k+n) \le \begin{bmatrix} -l_2 \\ 0 \end{bmatrix} for\!: k = N_p \quad (18\text{-}2)$$

For the CDHVs in the "far" location in the preceding position of the CHDV in "near" location on the target lane, the constraints, which are different from those above, are:

$$\hat{x}(k+n) - \bar{x}(k+n) \ge \begin{bmatrix} l_2 \\ -\hat{\delta}(k+n-1) \end{bmatrix} for\!: k = 1, \ldots, N_c \quad (19\text{-}1)$$

$$\hat{x}(k+n) - \bar{x}(k+n) \ge \begin{bmatrix} -l_2 \\ 0 \end{bmatrix} for\!: k = N_p \quad (19\text{-}2)$$

The constraints (18-1), (18-2), (19-1), (19-2) are applied to promise the safe longitudinal distance between the neighboring CHDVs in different locations.

## IV. SIMULATIONS AND RESULTS

This study developed simulation under different sets of conditions. The key points considered in the simulations are:

- The initial state of the CHDV on the target lane; this consists of the initial longitudinal location and initial velocity.
- The cooperative levels of the CHDV, which can be categorized as active cooperation and inactive cooperation.
- Multiple lane-changing simulations; therefore, more than one lane-change maneuver, is considered.

Based on these considerations, there exist several different combinations of initial state, in the simulations. The combinations can be placed into different categories in terms of percentage of actively cooperate CHDVs (Figure 6). The first refers to the 0% of actively cooperate CHDVs; the second refers to 50% of actively cooperate CHDVs; the third refers to 100% of actively cooperate CHDVs. In this section, the influence of the cooperative level and location is discussed in detail. Also, the type that has larger influence on the system efficiency serve as the focus in the simulation in terms of feasibility, and overall lane-changing time.

$d_{\max}$ = -5.08 m/s$^2$; maximum acceleration rate, $a_{max}$ = 5.08 m/s$^2$; maximum longitudinal acceleration rate through lane-changing maneuver, $a^L_{max}$ = 3.024 m/s$^2$; time interval for each time step, $\tau$ = 0.2s; reference TCAV trajectory and velocity r; longitudinal safety distance requirement, $l_1$ = 5m; longitudinal safety distance requirement, $l_2$ = 10m; $l_2$; vehicle length, $l_v$ = 4m; radius of vehicles' buffer circle, $l_r$ = 3m; lane width, $l_w$ 3.7m; weight parameter, $P$ = 15; weight parameter, $Q$ = 10; weight parameter, $R$ = 10; prediction horizon, $N_p$ = 5; control horizon $N_c$ = 4. The prediction horizon $N_p = 5$, which is based on sufficient condition for stability that mentioned in our previous study. -5.08 is the maximum longitudinal deceleration of the human driver to prevent an emergency situation (Bae et al., 2019). $P, Q, R$ are defined appropriately to have the objective function as a convex function. In order to have a dense traffic on the target lane, four vehicles are considered in all simulations. Two vehicles in the "near" location, and the other two vehicles in the "far" location. Also, their initial location is decided by the bumper-to-bumper distance based on their initial velocities. The vehicle length is 4m, the radius of the circle buffer area is defined as 3m (the diameter of the circle buffer area is 6m).

### Percentage of actively cooperate CHDVs

As mentioned previously, the CHDV at different locations can choose their cooperate level. When there are different number of actively cooperate CHDVs in the system, the performance of the control framework is affected. The actively cooperate CHDVs are in the different location in terms of "near" and "far". For the various combinations of actively CHDVs in the different locations, the optimization problem will differ. In this section, the performance of the control framework is evaluated in two different matrices: (a). Initial feasibility under different situation in terms of the actively cooperate CHDVs. (b). Multiple lane-changing process time under different combinations.

In Figure 6, there are three different combinations of actively cooperate CHDVs. The first sub-figure shows the percentage of actively cooperate CHDVs is the smallest, since all the CHDVs in the system are not actively cooperate. The second sub-figure shows the actively cooperating percentage is 50%. However, for the 50% actively cooperate CHDV, there are several different combinations in terms of the location of the CHDVs. The last sub-figure has the highest percentage active cooperation (100%).

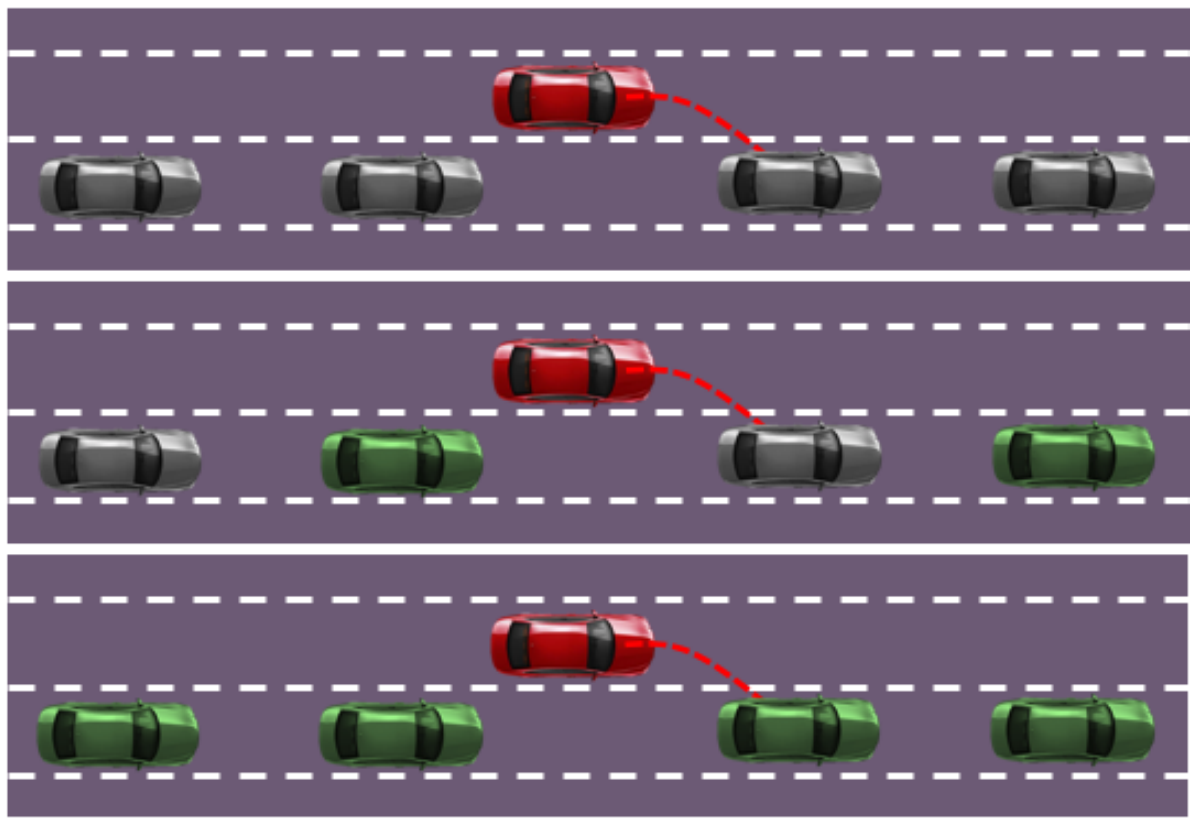

**FIGURE 6 Different combinations of actively cooperate CHDVs**

**Feasibility of different percentage of actively cooperate CHDVs**

The control framework's performance is affected by different combinations of the percentage of actively cooperate CHDVs. Feasibility is one of the important factors that needs to be considered. According to previous research, lane change is the main driving context that influences the efficiency of freeway traffic operations, particularly at locations near the ramp area. For example, when the lane-changing process is not feasible, the CAV will not be able to lane change directly, which means the CAV needs to wait for an available gap before it can carry out the lane-changing maneuver. This will cause the low efficiency or even congestion of the traffic flow behind the CAV. However, if the feasibility can be satisfied, the CAV can make lane change immediately without affecting significantly, the efficiency of other vehicles.

The cooperative control framework gives the infeasible situation a second chance by adjusting the CHDVs' acceleration/deceleration rates. In the meantime, the CAV will keep waiting for the CHDVs by maintaining the original speed. The initial feasibility will reach a higher level because of the "waiting-adjusting" strategy. Moreover, when multiple lane change is considered, the connection between two lane-change processes will need to consider the "waiting-adjusting" strategy because the feasibility of the second lane-change process is uncertain.

Figure 7 shows four sub-figures under the different combinations. The first sub-figure shows the actively cooperate CHDVs at 0%, the feasibility rates are low particularly when the standard deviations of the velocities are high. The second and third sub-figures have the actively cooperate CHDVs at 50% but in different locations of the CAV. Here, we test the following position and preceding position. The feasibility rates are improved under both situations. However, there are slightly differences: the preceding position seems to have larger effects on the feasibility rates. The last sub-figure is when actively cooperating CHDV is at 100%, and it is observed that the feasibility rates improved very significantly. There is a common pattern in all the figures: the larger the standard deviation, the lower the feasibility rate. However, the change of the average speed of all the vehicles seems have less influence on feasibility. Thus, we can tell that the relative velocities among the vehicles is an important factor of the system feasibility.

Furthermore, the feasibilities under different cooperative combinations are different. When the percentage of actively cooperating CHDVs is the highest, the average feasibility is the highest (that is, 0.94). This is because active cooperation gives a higher chance for the CAV to maintain at the current state and wait for a feasible situation that create by the surrounding

CHDVs. The feasibility when there is no actively cooperating CHDVs, is 0.35. In the situations where the percentage of actively cooperating CHDVs is 50%, the average feasibility rate is approximately 0.68, which still represents an improvement compared to the situation where there is 0% actively cooperating CHDVs.

Under the situations where the FHDV is feasible but the PHDV is not feasible, the actively cooperate PHDV will be easy to adjust to be feasible, to reach a feasibility rate of as much as 77%. However, in situations where PHDV is feasible and FHDV is not feasible, the actively cooperate FHDV will not be able to easily adjust the situation to feasible. Also, the feasible rate is 59%. Thus, in the situation where the FHDV is not feasible, it is more difficult for the CAV to start changing lanes.

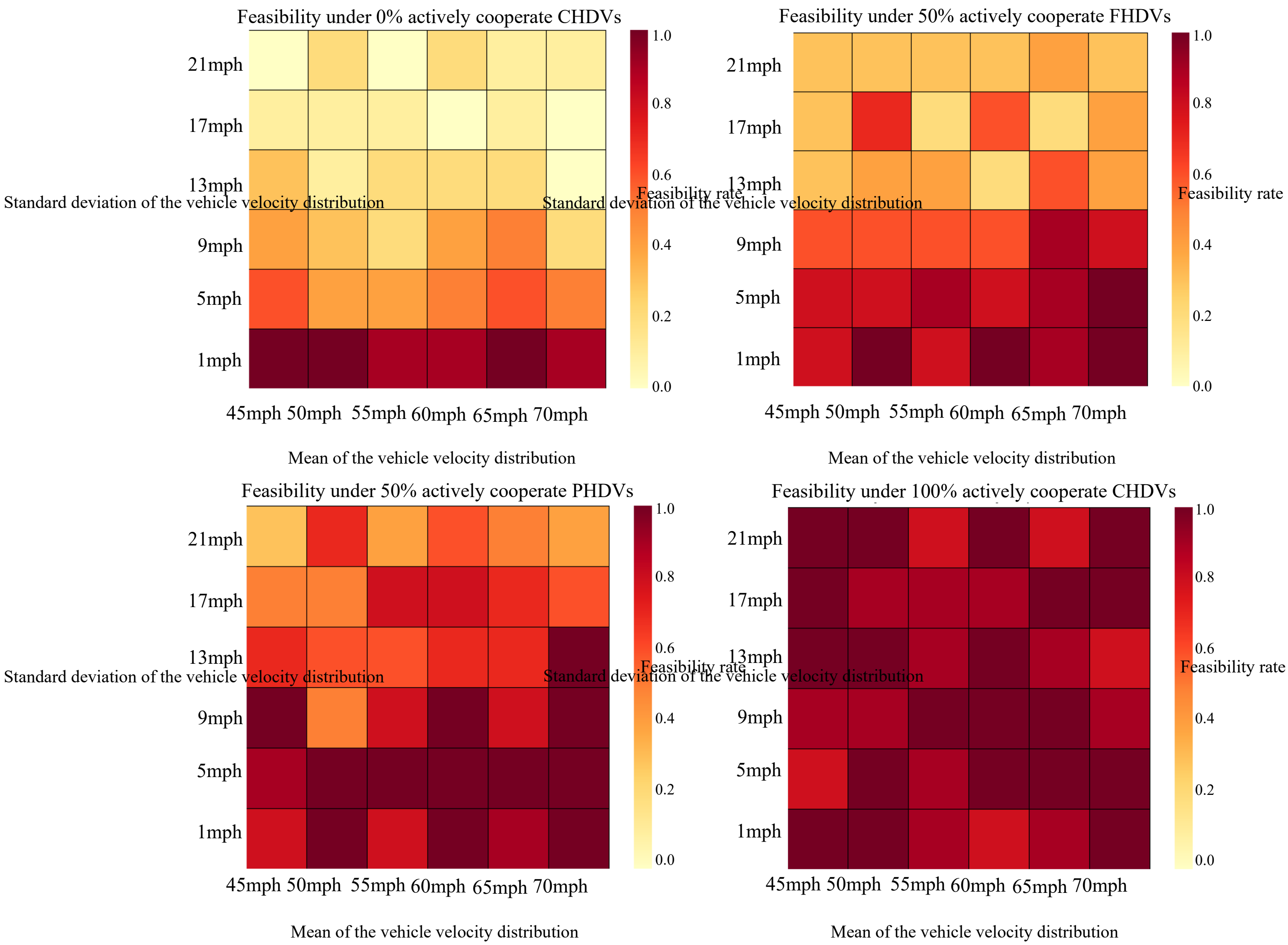

**FIGURE 7. Feasibility of four different cooperative percentage**

**Multiple lane-changing process time under different situations**

Efficiency is an important evaluation metric of the control framework performance. While feasibility is a key factor in evaluating different cooperative combinations, it cannot reflect the relations between the different combination and lane-changing efficiency. A straightforward way is to analyze the total lane-changing process time. In this research, we focused on multiple lane-changing maneuvers. Thus, the simulations in this section are bi-lane changes (Figure 8).

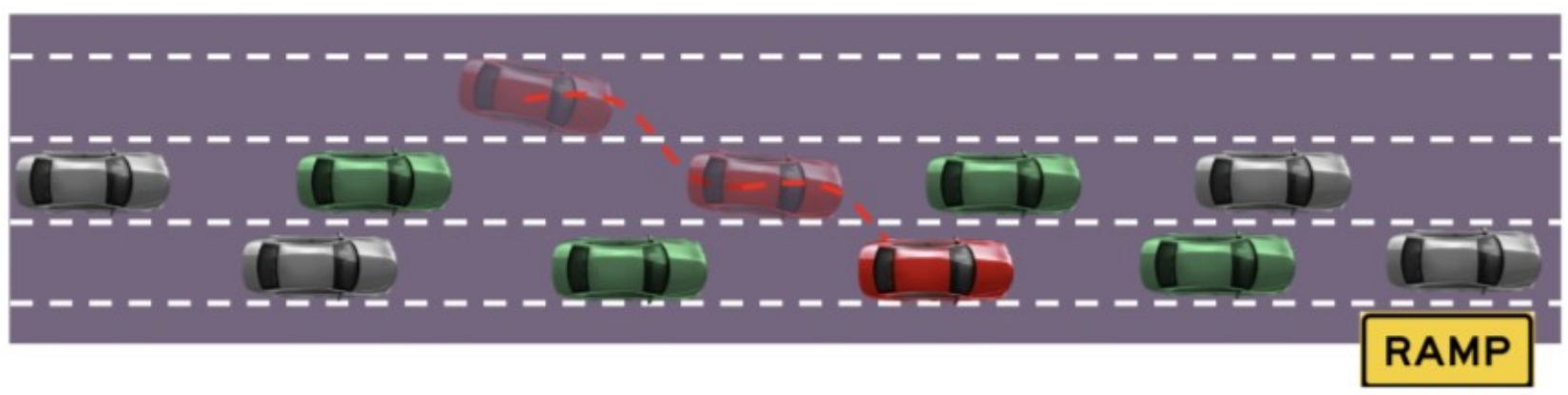

**FIGURE 8. Bi-lane change process**

The total lane-changing process time is the sum of the two lane-changing maneuvers' time and the possible waiting time before each lane-changing maneuver starts. Since there are two different lane-changing maneuvers that need to be considered, the number of combinations under multiple lane-changing maneuvers is doubled. Figure 9 shows the total lane-changing process time with different percentage of actively cooperating CHDVs. Considering the bi-lane changes, we have four different combinations, which are 0% of actively cooperating CHDVs in both lane-changing process, 50% of actively cooperating CHDVs in both lane-changing process, 50% of actively cooperating CHDVs in one lane-changing process, and 100% of actively cooperate CHDVs in the other, 100% of actively cooperating CHDVs in both lane-changing process. In the simulation, the actively cooperating CHDVs will have larger rates of acceleration/deceleration compared to 0% actively cooperating CHDV. Moreover, the waiting time for the CAV is set to be 2 seconds, which is 20-time steps.

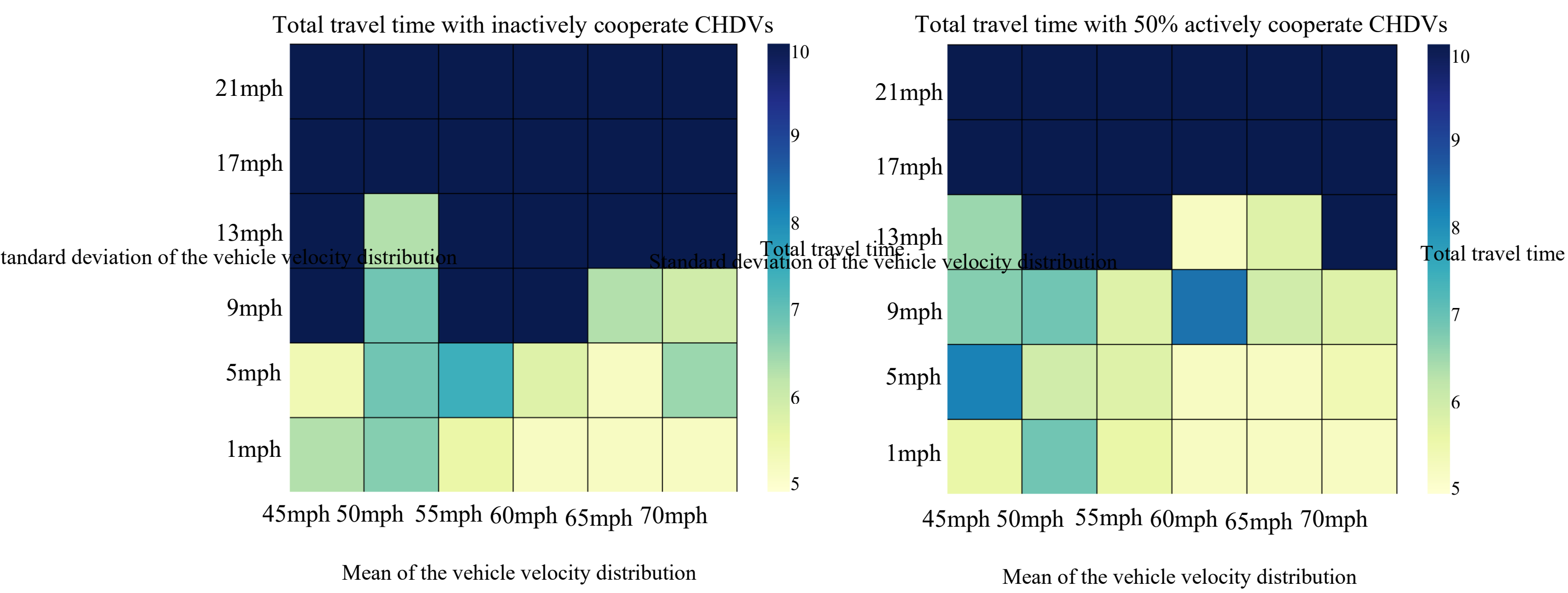

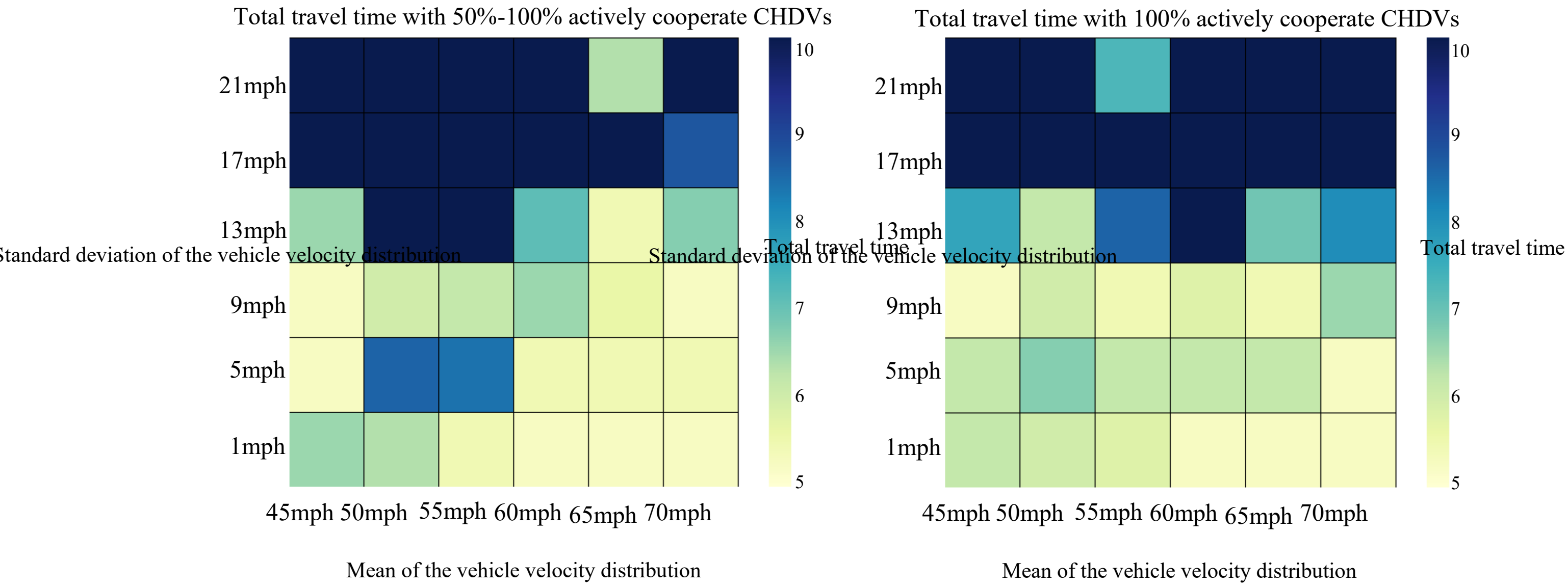

**FIGURE 9. Total travel time under different percentage of actively cooperate CHDVs**

In reality, when the mandatory lane change is caused by emergency situation, such as sudden lane drop, construction work zone, or queue-jumping (Zhang et al., 2018) to veer off the road, a smaller waiting time need to be considered. Thus, the framework will be more flexible to deal with emergency situations. With regard to normal conditions, the waiting time can be longer, the longer waiting time for CAV is set to be 6 seconds.

In Figure 9, the CAV waiting time is set to be smaller than 6 seconds. Normally, the lane-changing process time for an HDV will be 5~6 seconds. Since here we consider bi-lane change process, thus, for the cases where the total lane-changing process time is larger than 10 seconds are not acceptable. As shown in the first sub-figure of Figure 9, when there are 0% actively cooperating CHDVs in the system for both lane-changing maneuvers, there are 56% of the cases have total lane-changing process time as more than 10 seconds, which are shown as the dark area. When the actively cooperating CHDVs' percentage improves to 50%. In the second sub-figure, the cases that have total lane-changing as more than 10 seconds lower to 42%, which means 58% of the cases can finish the bi-lane change within 10 seconds. When one of the lane-changing process has 50% of actively cooperating CHDVs and the other has 100% of actively cooperating CHDVs, the cases of short lane change time improves to 67%. The higher the percentage of the actively cooperating CHDVs, the larger the "high-efficiency" area (the light-colored grids). However, when the percentage of actively cooperating CHDVs get 100% for both lane-changing maneuver, it is difficult to get further improvements of the control performance. Since the standard deviation of the dark area are too large, it is difficult for the control framework to yield further improvements. Also, there are common patterns that can be observed across the results for the different combinations: the higher the standard deviation, the lower the efficiency. The highest efficiency focus on the initial condition where the velocity distribution has small standard deviation, and the range of the mean velocity is 55mph to 70mph.

Furthermore, the location of the actively cooperate CHDVs will also affect the performance of the control framework. In Figure 10, the left sub-figure shows the situation where actively cooperate CHDVs are in the "near" location of the CAV, the sub-figure on the right shows the situation where actively cooperate CHDVs are in the "far" location of the CAV. The efficiency (total travel time) of the left sub-figure is apparently higher than the right subfigure. The cases of efficient lane-changing maneuvers take 67% when the actively cooperate CHDVs are in the "near" location. But the efficient lane-changing maneuvers take 56% when the actively cooperate CHDVs are in the "far" location.

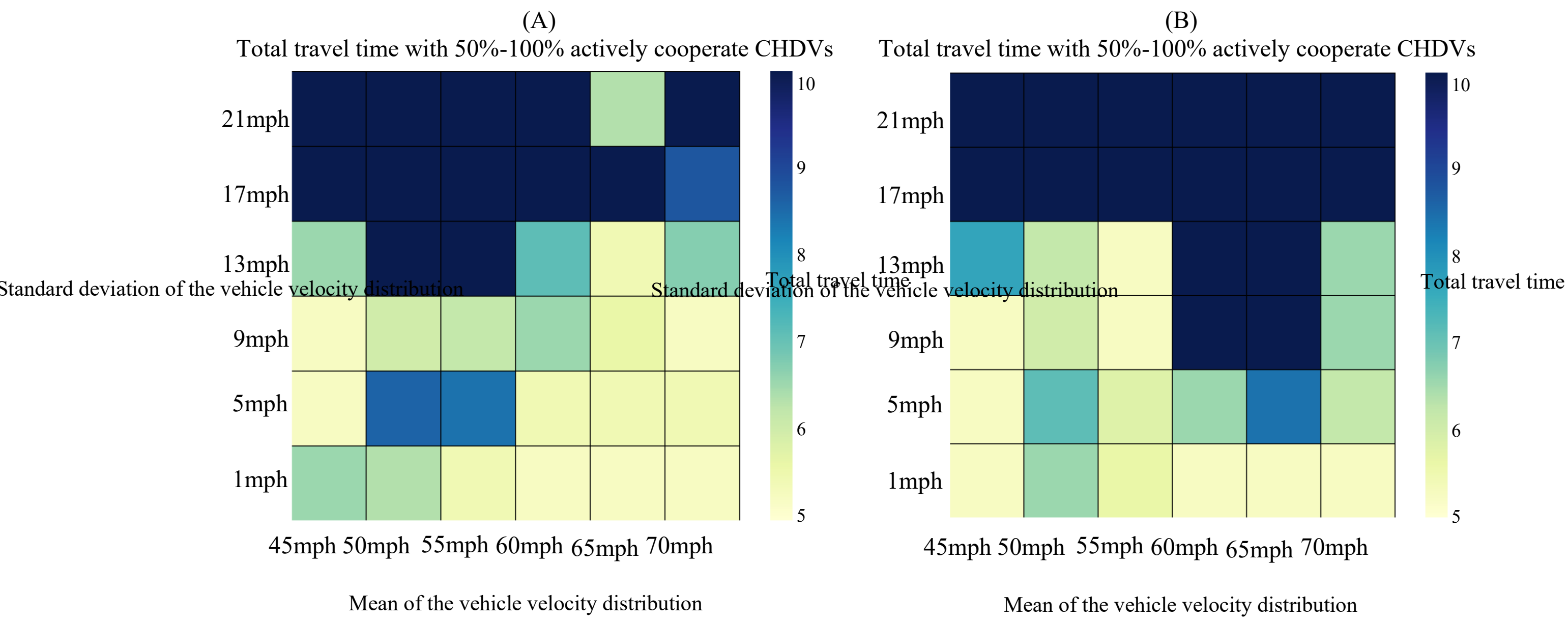

(a)Actively cooperating CHDVs in "near" location (b) Actively cooperating CHDVs in "far" location

**FIGURE 10. actively cooperate CHDVs in different location**

## V. CONCLUSIONS

This paper focused on the challenges that AV will face in the mixed flow era. Lane-changing maneuver under high traveling speed situations are risky and have low efficiency according to previous studies on the HDV lane-changing maneuvers. However, in the mixed flow era, it will be much more difficult for the AV to carry out this maneuver because the AVs' on the road are required to be "sufficiently" safe and it is expected that most AVs will be programmed and controlled to behave conservatively. Thus, this paper formulated a cooperative control framework for connected HDVs to cooperate with the CAV to help them carry out safe and efficient lane-changing maneuvers. The cooperative control framework is based on the vehicle automation and V2V connectivity. The underlying theory of the framework, MPC control, is used to address the multi-car interactions and multiple constraints in the CHDVs optimal motion planning. One important factor in the cooperative control framework is the level at which the human driven vehicles cooperate with the CAVs, and their different initial locations just before the CAV's prospective lane change. Since CHDVs are HDVs with connectivity, human drivers' compliance will affect the performance of the control framework. Thus, considering the cooperative level of different CHDVs is truly important. In the control framework design, we use two different levels to represent the drivers' compliance: inactive cooperation and active cooperation. We design the optimization problem in the MPC controller based on the cooperative level as well as the initial locations of the vehicles. We also assume that the human driver in the CHDVs will not accept a certain threshold level of sharp change in speed. Even though we have this severe penalty for the passenger comfort of the controlled CHDVs is represented by the jerk cost function in the optimization problem.

In order to test the framework performance under different combinations of the level of cooperation. We generate these different combinations in terms of the percentage of actively cooperate CHDVs in the system, and we considered different percentage levels: 0%, 50% and 100%. Moreover, various of initial states of the system are considered in the simulations. We consider different initial states based on velocity distributions in terms of velocity mean and standard deviation.

In the simulation part, the performance of the cooperative control framework is considered in two perspectives, the first is the initial feasibility under different percentages of actively cooperating CHDVs, and the other is the total process time of bi-lane change simulations. From the results, we observe that the higher the percentage of actively cooperating CHDVs in the system, the higher the feasibility rate. With regard to the lane-changing process duration, we observe that when the percentage of actively cooperating CHDVs is higher, the efficiency of those cases (which reflected by the total lane-changing process time) is higher. Clearly, the level of active cooperation of the CHDVs is really important, because it affect the performance of the control framework in a very significant way. We also observe that when the actively cooperate CHDVs in the different location, the performance is also different: when the actively cooperate CHDVs are in the "near" location of the CAV, the efficiency of lane change is higher. It was noted that the feasibility rates of starting lane-changing maneuvers are affected by the locations of the actively CHDVs under different situations. FHDV affects the feasibility rates larger than the PHDV, PHDV is easier to adjust by the control framework.

This research demonstrates the benefits of the HDV-CAV cooperative network to the CAV in the mixed traffic stream, and the importance of the cooperation (via compliance) of CHDVs in the neighborhood of the CAV. An actively cooperating vehicle system will yield high efficiency not only for CAV users but also for the overall system including the human drivers in CHDVs.

## ACKNOWLEDGMENTS

This work was supported by Purdue University's Center for Connected and Automated Transportation (CCAT), a part of the larger CCAT consortium, a USDOT Region 5 University Transportation Center funded by the U.S. Department of Transportation, Award #69A3551747105. The contents of this paper reflect the views of the authors, who are responsible for the facts and the accuracy of the data presented herein, and do not necessarily reflect the official views or policies of the sponsoring organization.

## AUTHOR CONTRIBUTIONS

The authors confirm contribution to the paper as follows: all authors contributed to all sections. All authors reviewed the results and approved the final version of the manuscript.